\documentclass[a4paper,14pt]{extarticle}
\usepackage{cmap} 
\usepackage[cp1251]{inputenc}
\usepackage[english,russian]{babel}

\usepackage{braket,amssymb,amsmath,array,commath,mathtext,wrapfig,tikz}
\usepackage[arrows=pgf-filled,version=3]{mhchem}
\usepackage{bm}
\usepackage{amsthm,amsfonts,amscd}
\usepackage{graphicx}
\usepackage{cite}
\usepackage{lscape}
\usepackage{geometry}
\usepackage{indentfirst}
\usepackage{afterpage,soul,ulem,dcolumn,changebar,longtable,hhline,multirow,makecell}
\usepackage[small,bf]{caption}
\makeatletter
\renewcommand{\@biblabel}[1]{#1.\hfil} % С„РёСЏ РІ \makeatother
\begin{document}
\renewcommand{\refname}{References}o

	\begin{center}
\textit{Extraction Energy From Charged Vaidya Black Hole Via Penrose Process}
	\end{center}
	
	\begin{center}
\textbf{Vitalii Vertogradov}\\
	\end{center}

	\begin{center}
		Physics department, Herzen state Pedagogical University of Russia,
		
		48 Moika Emb., Saint Petersburg 191186, Russia
		
		SPb branch of SAO RAS, 65 Pulkovskoe Rd, Saint Petersburg 196140, Russia
		
		vdvertogradov@gmail.com
	\end{center}

\textbf{Summary}. In this paper, we consider the analogy of the Penrose process in Charged Vaidya spacetime. We calculate the border of the generalized ergosphere, in which the charged particles with negative energy might exist,  and show that it is temporary. We show that there are no closed orbits for particles with negative energy inside the generalized ergosphere. We investigate the question  about the efficiency of the Penrose process and show that one can't extract large energies from a black hole.


\section{Introduction}

Black holes play major role in modern theoretical physics. The direct observations~\cite{bib:bhob1, bib:bhob2, bib:bhob3} shows that a black hole is a real astrophysical object. 

There are several ways how the energy can be extracted from a black hole~\cite{bib:pen, bib:14, bib:15}. Penrose process states that in the ergoregion of a rotating black hole the particles with negative energy might exist due to collision or decay. As the result the energy of escaping particle is bigger than the energy of the original particle. This process is possible only in the ergorigion where the Killing vector $\frac{d}{dt}$ becomes spacelike. In spherical symmetry, i.e. for Schwarzschild and Reissner-Nordstrom black holes the Killing vector $\frac{d}{dt}$ is timelike outside the event horizon and the Penrose process is impossible. However, in Reissner-Nordstrom case, one can consider the analogy of the Penrose effect because there are charged particles with negative energy~\cite{bib:Chandra}. These particles can exist only in the generalized ergoregion~\cite{bib:zaslav}. The border of this region is called generalized ergosphere and it depends on the test particle parameters. Thus, even in the spherically-symmetric case there is the analogue of the Penrose process for charged black holes~\cite{bib:zaslav, bib:papapetroo}. 

In the present work, we consider dynamical charged black hole which exterior geometry is described by charged Vaidya solution~\cite{bib:charg1, bib:charg2}. The Vaidya metric describes a dynamical spacetime instead of a static spacetime as the Schwarzschild or Reissner-Nordstrom  metrics do. In the real world, astronomical bodies gain mass when they absorb radiation and they lose mass when they emit radiation, which means that the space-time around them is time-dependent.  In dynamical spherically-symmetric spacetimes, in general, there is only one conserved quantity - angular momentum $L$. However, for certain choice of the mass $M(v)$ and charge $Q(v)$ functions, there are additional symmetry due to a conformal Killing vector field. We calculate the location of the generalized ergosphere where the charged particles with negative energy can exist. After that, we consider the Penrose effect and estimate the efficiency of this process.

This paper is organized as follows. In sec. II we consider the charged Vaidya spacetime, transform it to conformally static coordinates and calculate the location of generalized ergosphere. In sec. III we show that the Penrose process can occur in the generalized ergoregion and in sec. IV we estimate the efficiency of the process. Sec. V is the conclusion.
The system of units $G=c=1$ will be used throughout the paper.

\section{Charged Vaidya spacetime}

The charged Vaidya spacetime in advanced EddingtonвЂ“Finkelstein coordinates has the following form:
\begin{equation} \label{eq:metric}
ds^2=-\left(1-\frac{2M(v)}{r}+\frac{Q^2(v)}{r^2} \right )+2dvdr+r^2d\Omega^2 \,.
\end{equation}

Here $M(v)$ and $Q(v)$ are mass and charge function which depend upon the time $v$, $d\Omega^2$ is the metric on unit two sphere.

The metric \eqref{eq:metric} violates the weak energy condition~\cite{bib:pois, bib:hok} at the region $r<r_C=\frac{Q\dot{Q}}{\dot{M}}$ (The overdot means the partial derivative with respect to time $v$ -$\dot{M}=\frac{\partial M}{\partial v}$.). However, particles can't across this boundary due to the Lorentz force~\cite{bib:lorentz}. We restrict the consideration by assumption that $r>r_C$. If we put $M(v)\equiv M=const.$ and $Q(v)\equiv Q=const.$ then the metric \eqref{eq:metric} becomes Reissner-Nordstrom in advanced Eddington-Finkelstein coordinates. If we consider neutral particles then there are no particles with negative energy outside the event horizon $r=r_h=M+\sqrt{M^2-Q^2}$. The same statement is valid in the metric \eqref{eq:metric}~\cite{bib:ver1}. The situation changes when one considers the non-geodesic motion of the charged particles~\cite{bib:Chandra, bib:zaslav}. 

The metric \eqref{eq:metric} is spherically symmetric and we can consider the motion in the equatorial plane $\theta=\frac{\pi}{2}$. From the spherical symmetry we have the constant of the motion  - the angular momentum $L$:
\begin{equation}
L=r^2\frac{d\varphi}{d\lambda} \,.
\end{equation}

Where $\lambda$ is the affine parameter.

The energy expression for the particle with an electric charge $Q^*$ reads:
\begin{equation}
E(v)=\left(1-\frac{2M}{r}+\frac{Q^2}{r^2} \right ) \frac{dv}{d\lambda}-\frac{dr}{d\lambda}+\frac{QQ^*}{r} \,.
\end{equation}

And using the fact that $g_{ik}u^iu^k=-1$ we formally obtain:
\begin{equation} \label{eq:wrong}
\left(\frac{dr}{d\lambda}\right)^2=\left(E(v)-\frac{QQ^*}{r}\right)^2-\left(1-\frac{2M}{r}+\frac{QQ^*}{r^2}\right) \left(\frac{L^2}{r^2}+1\right) \,.
\end{equation}

Now, one should  express the energy to understand the possible minimum of energy.
\begin{equation}
E_{max}=\frac{QQ^*}{r}+\sqrt{\left( 1-\frac{2M}{r}+\frac{QQ^*}{r^2}\right)\left(\frac{L^2}{r^2}+1\right)+\left(\frac{dr}{d\lambda} \right)^2} \,.
\end{equation}
We  chose the 'plus' sign due to the fact that the energy $E$ must be positive for particles  at infinity in the case $Q^*=0$. One can have the negative energy $E$ only in the case $Q^*<0$ i.e. a test particle has the opposite electric charge than a black hole. The energy $E$ has its minimum if we put $\frac{dr}{d\lambda}=0$ and  $L=0$, substituting these conditions into the equation above, one can obtain the generalized ergosphere:
\begin{equation} \label{eq:generalized}
-QQ^*\geq r^2\sqrt{1-\frac{2M}{r}+\frac{QQ^*}{r^2}} \,.
\end{equation}
 One should understand that we can't take the charge of a black hole $Q$ as large as one wants due to the fact that we have condition of black hole formation $M\geq Q$. Otherwise, one has the naked singularity which we don't consider here. 

However, the metric \eqref{eq:metric} is time-depended and as the result the energy is not a constant. So, to investigate the Penrose process, the equation \eqref{eq:wrong} doesn't suit us.

In the general case, the  charged Vaidya spacetime doesn't possess any additional symmetry. However, for the particular choice of the mass and charge function, the metric \eqref{eq:metric} admits the conformal Killing vector~\cite{bib:nel}. In this case $M$ and $Q$ must  have the following form:
\begin{equation} \label{eq:spec}
\begin{split}
M(v)=\nu v \,, \nu>0 \,, \\
Q(v)=\mu v \,, \mu\neq 0 \,.
\end{split}
\end{equation}

To avoid the naked singularity, one should demand $\nu \geq \mu$. Thus, by doing the following coordinate transformation~\cite{bib:germany}:
\begin{equation}
\begin{split}
v=r_{0} e^{\frac{t}{r_0}} \,, \\
r=R e^{\frac{t}{r_0}} \,.
\end{split}
\end{equation}

one obtains the charged Vaidya spacetime in conformally static coordinates:
\begin{equation}
ds^2=e^{\frac{2t}{r_0}}\left [ -\left(1-\frac{2\nu r_0}{R}+\frac{\mu^2r_0^2}{R^2}-2\frac{R}{r_0}\right)dt^2+2dtdR+R^2d\Omega^2\right] \,.
\end{equation}

The conformal Killing vector becomes:
\begin{equation} \label{eq:killingconformal}
v\frac{d}{dv}+r\frac{d}{dr}=r_0\frac{d}{dt} \,.
\end{equation}
And it is timelike when :
\begin{equation} \label{eq:timelikecondition}
1-\frac{2\nu r_0}{R}+\frac{\mu^2r_0^2}{R^2}-2\frac{R}{r_0} >0 \,.
\end{equation}

Now, we have two constants of motion - energy$E$ and angular momentum $L$ :
\begin{equation} \label{eq:energy}
\begin{split}
E=e^{\frac{2t}{r_0}}\left [\left( 1-\frac{2\nu r_0}{R}+\frac{\mu^2r_0^2}{R^2}-2\frac{R}{r_0} \right) \frac{dt}{d\lambda}-\frac{dR}{d\lambda}\right]+\frac{\mu Q^*r_0}{R} \,, \\
L=e^{\frac{2t}{r_0}} R^2\frac{d\varphi}{d\lambda} \,.
\end{split}
\end{equation}

Substituting \eqref{eq:energy} into the normalization condition $g_{ik}u^iu^k=-1$ we obtain:
\begin{equation}
e^{\frac{4t}{r_0}}\left( \frac{dR}{d\lambda} \right )^2=\left(E-\frac{\mu Q^* r_0}{R}\right)^2-e^{\frac{2t}{r_0}}\left( 1-\frac{2\nu r_0}{R}+\frac{\mu^2r_0^2}{R^2}-2\frac{R}{r_0} \right) \left(\frac{L^2}{R^2Рµ^{\frac{2t}{r_0}}}+1\right) \,.
\end{equation}

Or, expressing the energy $E$, we have:
\begin{equation} \label{eq:energy2}
E=\frac{\mu Q^*r_0}{R}+e^\frac{t}{r_0}\sqrt{\left( 1-\frac{2\nu r_0}{R}+\frac{\mu^2r_0^2}{R^2}-2\frac{R}{r_0} \right)\left(\frac{L^2}{R^2 e^{\frac{2t}{r_0}}}+1\right)+ e^{\frac{2t}{r_0}}\left(\frac{dR}{d\lambda} \right)^2} \,.
\end{equation}
 One should  pick up the plus sign  like we did it above because the observer which moves along $t$ line must measure the positive energy for neutral particles at $R=r_0$. Again, we have the negative energy only if $Q^*<0$ i.e. the particle and a black hole have opposite electric charges. To obtain the generalized ergosphere one should put $\frac{dR}{d\lambda}=0$ and $L=0$ in \eqref{eq:energy2}. By doing this we obtain the region where particles with negative energy can exist:
\begin{equation} \label{eq:generalizedergosphere}
-\mu Q^* r_0\geq e^{\frac{t}{r_0}} R\sqrt{1-\frac{2\nu r_0}{R}+\frac{\mu^2r_0^2}{R^2}-2\frac{R}{r_0} } \,.
\end{equation}
The extremal case of the inequality \eqref{eq:generalizedergosphere} gives the border of the generalized ergosphere. The first thing which one can notice from \eqref{eq:generalizedergosphere} that the generalized ergosphere is temporal and its border is shrinking during the evolution of the system. It means that the charged particles with negative energy can exist only the same period of the time. Hence, all such particles must inevitably  fall on the apparent horizon and closed orbits can't exist for such particles because as soon  the generalized ergosphere disappears as only particles with positive energy are allowed. 

Note that in Kerr metric the ergosphere is the surface on which the Killing vector $\frac{d}{dt}$ becomes null and, as the result, it is defined purely geometrically. The generalized ergosphere \eqref{eq:generalizedergosphere}, on the other hand, depends upon the particle parameters, especially upon the particle charge $Q^*$. As the result, for each particle the border of the generalized ergosphere and  the time of its existence are different. This quantity shows only the region where the particle with certain parameters might possess the negative energy.

\section{Penrose process}

In this section we investigate the energy extraction process which was proposed by Penrose~\cite{bib:pen}. The mechanism, which we use, are the following: a negatively charged particle gets into the generalized ergosphere, breaks up into two fragments, one of which escapes to the observer at $R=r_0$ ($r_0$ must be in the region where the conformal Killing vector $\frac{d}{dt}$ is timelike.) with more energy than the original particle. The remaining particle with negative energy falls onto a black hole which we showed in the previous section. Let's denote the initial particle with subscript $0$, outgoing particle by $2$ and falling particle as $1$. We consider the equatorial motion, so the break up point is located in $\{R^*\,, \varphi^*\}$ point which must be inside of the generalized ergosphere. 

The quantities that characterize each particle are related by conservation equations. Charge conservation, for instance, yields:
\begin{equation}
Q^*_0 m_0=Q^*_1 m_1+Q^*_2 m_2 \,.
\end{equation}

The conservation of the four-momentum reads:
\begin{equation} \label{eq:cmomentum}
p^\mu_0=p^\mu_1+p^\mu_2 \,.
\end{equation}
The equations \eqref{eq:cmomentum}  are three different conservation equations with straightforward physical interpretation. The zero-component is the conservation of energy:
\begin{equation} \label{eq:cenergy}
E_0m_0=E_1m_1+E_2m_2 \,,
\end{equation}
the $R$ components corresponds the conservations of linear momenta:
\begin{equation} \label{eq:cmomenta}
m_0\left(\frac{dR}{d\lambda} \right)_0=m_1\left(\frac{dR}{d\lambda}\right)_1+m_2\left(\frac{dR}{d\lambda}\right)_2 \,,
\end{equation}
and $\varphi$-component means the angular momentum conservation:
\begin{equation}
m_0L_0=m_1L_1+m_2L_2 \,.
\end{equation}

Note that all derivatives in \eqref{eq:cmomenta} are evaluated at the break-up point. Also, if we square \eqref{eq:cmomentum}by  keeping in mind the fact that $|p^\mu_i|^2=m^2_i$ and $p^\mu$ iis timelike and future-directed, we come to the inequality:
\begin{equation} \label{eq:1mass}
m_0^2\geq m_1^2+m_2^2 \,.
\end{equation}
The energy which is carried away by the particle $2$ is, according to \eqref{eq:cenergy}:
\begin{equation} \label{eq:process}
m_2E_2=m_0E_0-E_1m_1 \,.
\end{equation}
As we found out in the previous section that a negatively charged particle in the generalized ergosphere and falling onto a black hole might possess the negative energy. If this is the case, then \eqref{eq:process} shows that $E_2>E_0$ and we have the extraction of the energy from the black hole. As we know that a particle $1$ with negative energy must fall onto a black hole, hence, it will directly decrease the energy associated with the black hole, as in Penrose's original proposal. 

\section{The maximum of the energy extraction}

The efficiency $\eta$ of the Penrose process can be defined as:
\begin{equation} \label{eq:etta}
\eta=\frac{m_2E_2}{m_0E_0}-1=-\frac{m_1E_1}{m_0E_0} \,.
\end{equation}
To maximize we should make $E_0$ as small as possible and $|E_1|$ as large as possible. In other words, we want to extract as much energy as possible starting with as little energy as possible. Let's put $E_0=1$ for simplicity. Also, we assume that a particle $0$ doesn't posses the angular momentum $L_0=0$. From \eqref{eq:energy2} the energy of a particle $1$ is most negative if the particle is initially at rest i.e.:
\begin{equation} \label{eq:energy3}
\begin{split}
\left(\frac{dR}{d\lambda}\right)_1=\left(\frac{d\varphi}{d\lambda}\right)_1=0 \,, \\
E=\frac{\mu Q^*r_0}{R}+e^\frac{t}{r_0}\sqrt{\left( 1-\frac{2\nu r_0}{R}+\frac{\mu^2r_0^2}{R^2}-2\frac{R}{r_0} \right)} \,.
\end{split}
\end{equation}

From the condition \eqref{eq:energy3} one can easily see that $L_0=L_1=L_2=0$. Now we should evaluate the mass $m_1$, for this purpose, let's express $m_2$ from \eqref{eq:cmomenta}, keeping in the mind that $\left( \frac{dR}{d\lambda}\right) _1=0$:
\begin{equation} \label{eq:mass2}
m_2^2=\left[ \frac{m_0\left( \frac{dR}{d\lambda}\right)_0}{\left(\frac{dR}{d\lambda}\right)_2}\right]^2 \,.
\end{equation}

Now, by using \eqref{eq:1mass}  and \eqref{eq:mass2}, we obtain the following restriction on $\eta$:
\begin{equation} \label{eq:maxenergy}
\eta < - \left[\frac{\mu Q^*r_0}{R}+e^\frac{t}{r_0}\sqrt{\left( 1-\frac{2\nu r_0}{R}+\frac{\mu^2r_0^2}{R^2}-2\frac{R}{r_0} \right)} \right]\sqrt{1-\frac{\left(\frac{dR}{d\lambda}\right)_0^2}{\left (\frac{dR}{d\lambda} \right)_2^2}} \,.
\end{equation}
From \eqref{eq:maxenergy} one can see that the maximum energy, which a particle $2$ can carry out, can't exceed the absolute value of the energy of the falling particle $1$. Like we have in the case of the boundary of generalized ergosphere, the efficiency $\eta$ depends on the particle charge $Q^*_1$. Note that if the radial velocities of falling and escaping particles are equal, then we don't have any energy extraction at all. To have energy extraction the following condition must be held:
\begin{equation}
\left|\left(\frac{dR}{d\lambda}\right)_2\right|>\left|\left(\frac{dR}{d\lambda}\right)_0\right| \,.
\end{equation}

The efficiency  $\eta$ is maximum if the collision takes place at the vicinity of $R=R_{max}$, where $R_{max}$ is the solution of the following equation:
\begin{equation} \label{eq:concol}
1-\frac{2\nu r_0}{R_{max}}+\frac{\mu^2r_0^2}{R^2_{max}}-2\frac{R_{max}}{r_0} =0 \,.
\end{equation}

It means that the collision takes place at the vicinity of the conformal Killing horizon. If we take into account the condition \eqref{eq:concol} then one obtains:
\begin{equation}
\eta<-\frac{\mu Q^*_1r_0}{R_{max}}\sqrt{1-\frac{\left(\frac{dR}{d\lambda}\right)_0^2}{\left (\frac{dR}{d\lambda} \right)_2^2}} \,.
\end{equation}

So a particle $2$ can carry away only a small amount of the energy. 

We found out that the charged particle with negative energy will fall onto a black hole. This particle must be opposite charged then a black hole. From this fact, we can conclude that The extraction energy decreases the energy associated with the black hole, charge. 

\section{Conclusion}

In this paper the analogue of the Penrose process has been considered in the case of charged Vaidya spacetime. Despite the fact that this spacetime is time-depended, it admits an additional symmetry due to a conformal Killing vector field for the special choice of the mass and charge functions. We have found out that the charged particles with negative can exist in the generalized ergoregion only if the an electric charge of the particle is opposite to the charge of a black hole. We have calculated the border of this region - generalized ergosphere and showed that its location depends upon the particle's charge. This border is shrinking. It means that the generalized ergosphere is temporary and it will disappear. Hence, closed orbits for charged particles with negative energy are impossible.

However, when the generalized ergosphere exist there is a possibility of the energy extraction from a black hole due to Penrose process. Since the charged particle must possess the opposite electric charge than a black hole one, the energy extraction is possible due to reducing a black hole charge. We have estimated the efficiency of this process and found out that the energy extraction is possible if the speed of the escaping particle is bigger than the speed of original one.  The maximum efficiency can be achieved only if the process occurs in the vicinity of the conformal Killing horizon. The energy which  is carried away  by the escaping particle is bounded and the efficiency can't exceed the absolute value of the energy of the falling particle.

\textbf{Acknowledgments} Author says thank to grant в„–В 22-22-00112В RSF for financial support and A. Samsonova for support.

\end{document}